\documentclass[10pt,superscriptaddress,twocolumn,amsmath,amssymb,aps,pra,showpacs]{revtex4-1}
\usepackage{mathrsfs}
\usepackage{graphicx}
\usepackage{dcolumn}
\usepackage{bm}
\usepackage[title,titletoc,header]{appendix}

\usepackage{amssymb}
\usepackage{amsmath}
\usepackage{mathrsfs}
\usepackage{amsmath}
\usepackage{subfigure}
\usepackage{float}
\usepackage[title,titletoc,header]{appendix}
\usepackage{graphicx}

\newcommand{\be}{\begin{equation}}
\newcommand{\ee}{\end{equation}}
\newcommand{\bea}{\begin{eqnarray}}
\newcommand{\eea}{\end{eqnarray}}

\begin{document}
\title{The odd-even effect of mosaic modulation period of quasi-periodic hopping  on the Anderson localization  in a one-dimensional lattice model  }
\author{Yi-Cai Zhang}\thanks{Corresponding author. E-mail:zhangyicai123456@163.com}
\affiliation{School of Physics and Materials Science, Guangzhou University, Guangzhou 510006, China}

\author{Rong Yuan}
\affiliation{School of Mathematical Sciences, Laboratory of Mathematics and Complex Systems,
MOE, Beijing Normal University, Beijing 100875, China}

\author{Shuwei Song}
\affiliation{Heilongjiang Provincial Key Laboratory of Quantum Control,
Harbin University of Science and Technology,
Harbin 150080, P. R. China}

\author{Mingpeng Hu}
\affiliation{School of Mathematical Sciences, Laboratory of Mathematics and Complex Systems,
MOE, Beijing Normal University, Beijing 100875, China}

\author{Chaofei Liu}
\affiliation{School of Science, Jiangxi University of Science and Technology, Ganzhou 341000, China}

\author{Yongjian Wang}\thanks{Corresponding author. E-mail:wangyongjian@amss.ac.cn}
\affiliation{School of Mathematics and Statistics Nanjing University of Science and Technology, Nanjing 210094, China}


\date{\today}
\begin{abstract}
In this study, we investigate Anderson localization in a one-dimensional lattice with a mosaic off-diagonal quasiperiodic hopping. Our findings reveal that the localization behavior of zero-energy states is highly dependent on the parity of the mosaic modulation period, denoted as $\kappa$. Specifically, when $\kappa$ is an odd integer, there is no Anderson localization transition even for large quasiperiodic hopping strengths, and the zero-energy state remains in a critical state. On the other hand, for an even $\kappa$ and a generic quasiperiodic hopping, the zero-energy state becomes a localized edge state at either the left or right end of the system. Additionally, we observe that the geometric mean value of the energy spectrum is equal to the constant hopping for an even $\kappa$, while for an odd $\kappa$, it is equal to the geometric mean value of the hopping. This odd-even effect of the mosaic period also extends to other eigenstates near zero energy. More specifically, for an odd $\kappa$, there exists an energy window in which the eigenstates remain critical even for strong quasiperiodic hopping. In contrast, for an even $\kappa$, an Anderson localization transition occurs as the hopping strength increases. Furthermore, we are able to accurately determine the Lyapunov exponent $\gamma(E)$ and the mobility edges $E_c$. By analyzing the Lyapunov exponent, we identify critical regions in the hopping-energy parameter planes. Additionally, as the energy approaches the mobility edges, we observe a critical index of localization length of $\nu=1$. Finally, we demonstrate that different systems can be characterized by their Lyapunov exponent $\gamma(E)$ and Avila's acceleration $\omega(E)$.

\end{abstract}

\maketitle
\section{Introduction}
 In a conventional orthogonal class system, it is believed that even a weakly uncorrelated diagonal disorder in one and two dimensions \cite{Economou2006} can lead to Anderson localization \cite{Anderson1957}. In three dimensions, there is a mobility edge $E_c$ that separates localized states from extended states \cite{Economou1972}. As the eigenenergies approach the mobility edge $E_c$, the localization length of localized states diverges. Interestingly, in the presence of off-diagonal uncorrelated disorders, one-dimensional systems can exhibit a singular density of states near zero energy \cite{Dyson, Eggarter1978, Balents1997}, resulting in anomalous localization \cite{Theodorou1976, Antoaiou1977,Soukoulis} where the localization length is proportional to the square root of the system size \cite{Fleishman1977, Inui1992, Izrailev2012}. When the energy deviates from zero, the eigenstates are typically localized. However, if the off-diagonal disorder is correlated, the system can undergo a localized-extended transition \cite{Cheraghchi2005}. 
  In the presence of both diagonal and correlated off-diagonal disorders, one-dimensional systems can also exhibit extended states \cite{Zhangwei2004}.
Additionally, the correlation between diagonal and off-diagonal disorders can lead to the emergence of $E_c$ in one-dimensional lattice models \cite{Flores}. 
  The off-diagonal disorder model  have been realized experimentally \cite{Martin, Kraus, Xiao2021}.
  Recently, a mosaic lattice model with diagonal quasiperiodic disorder has been proposed \cite{Wangyucheng2020}. This model has been found to have mobility edges, which can be exactly determined using Avila's theory \cite{Liu2021, Avila2015}.

Since there exist above anomalous properties of  localizations in the  model with pure off-diagonal uncorrelated disorder,
a natural question arises, how is it if the off-diagonal hopping is quasiperiodic?
 One may wonder if there are mobility edges for off-diagonal quasiperiodic disorder (hopping). What are the localization properties of the eigenstates?

In this work, we try to answer the above questions by exploring a quasiperiodic off-diagonal disorder model with mosaic modulation.  The model is described by the following equation:
 \begin{align}\label{2}
V_{i,i+1}\psi(i+1)+V_{i,i-1}\psi(i-1)=E\psi(i).
\end{align}
where
 \begin{align}
V_{i,i+1}=V_{i+1,i}=
\left\{\begin{array}{cc}
t,  & for \ i\neq 0 \ mod \ \kappa\\
\frac{2\lambda \cos(2\pi \beta i+\phi)}{\sqrt{1-\tau \cos^2(2\pi \beta i+\phi)}},& for \ i= 0 \ mod \ \kappa
  \end{array}\right.
\end{align}
Here, $t>0$ represents the constant hopping strength, $\lambda$ describes the quasi-periodic hopping strength, $\kappa$ is a positive integer representing the mosaic period, $\beta$ is an irrational number, and $\tau$ is a real number. In this study, we only consider values of $\tau$ that are less than or equal to $1$, that is, $\tau\leq 1$ and integer $\kappa$ satisfies $\kappa\geq2$. Furthermore, it is important to mention that the above model (Eq. 1) does not have any extended states. This is due to the fact that, according to \cite{Barry1989} (or the mechanism described in \cite{xwzl}), the absolutely continuous spectrum, which corresponds to extended states, is empty because there exists a sequence $\{n_k\}$ such that $V_{n_k,n_{k}+1}\rightarrow0$. Therefore, any mobility edges (if present) would separate localized states from critical states. Throughout this paper, we have chosen $\beta=(\sqrt{5}-1)/2$ and used the units of $t=1$.

The parity of the mosaic period $\kappa$ has been found to have a significant impact on the localization of eigenstates near zero energy. Specifically, if the mosaic period $\kappa$ is odd, there is no Anderson localization for arbitrarily strong  quasi-periodic hopping strength $\lambda$. However, for an even integer mosaic period, the system undergoes Anderson localization as the quasi-periodic hopping increases. Furthermore, the Lyapunov exponent $\gamma(E)$ and mobility edges can be accurately determined using the Avila's theory.
With Lyapunov exponent, we find that some critical regions in the parameter plane would appear.
 In comparison with the localized states, the spatial extensions of eigenstates and their fluctuations in the critical region are much larger.
Near localized-critical transition points ($E_c$),  the localization length diverges, i.e.,
\begin{align}\label{10}
\xi(E)\equiv1/\gamma(E)\propto|E-E_c|^{-\nu}\rightarrow\infty, \ \ as \ E\rightarrow E_c,
\end{align}
where the critical index \cite{Huckestein1990}  is $\nu=1$, which differs from the $\nu=2$ of the correlated diagonal and off-diagonal disorder model \cite{Flores}.
Finally, we show that systems with different parameters $E$ can be systematically classified by the Lyapunov exponent and the Avila acceleration.

The work is organized as follows. First of all, we discuss the localization properties of zero-energy states for both odd and even number $\kappa$ in Sec.\textbf{II}.  In Sec.\textbf{III}, the Lyapunov exponent is calculated.  Next, with the Lyapunov exponent, we  determine the mobility edges and critical region  in Sec.\textbf{IV}. In addition, Avila's acceleration is also calculated.
 At the end, a summary is given in Sec.\textbf{V}.

\begin{figure}
\begin{center}
\includegraphics[width=1.0\columnwidth]{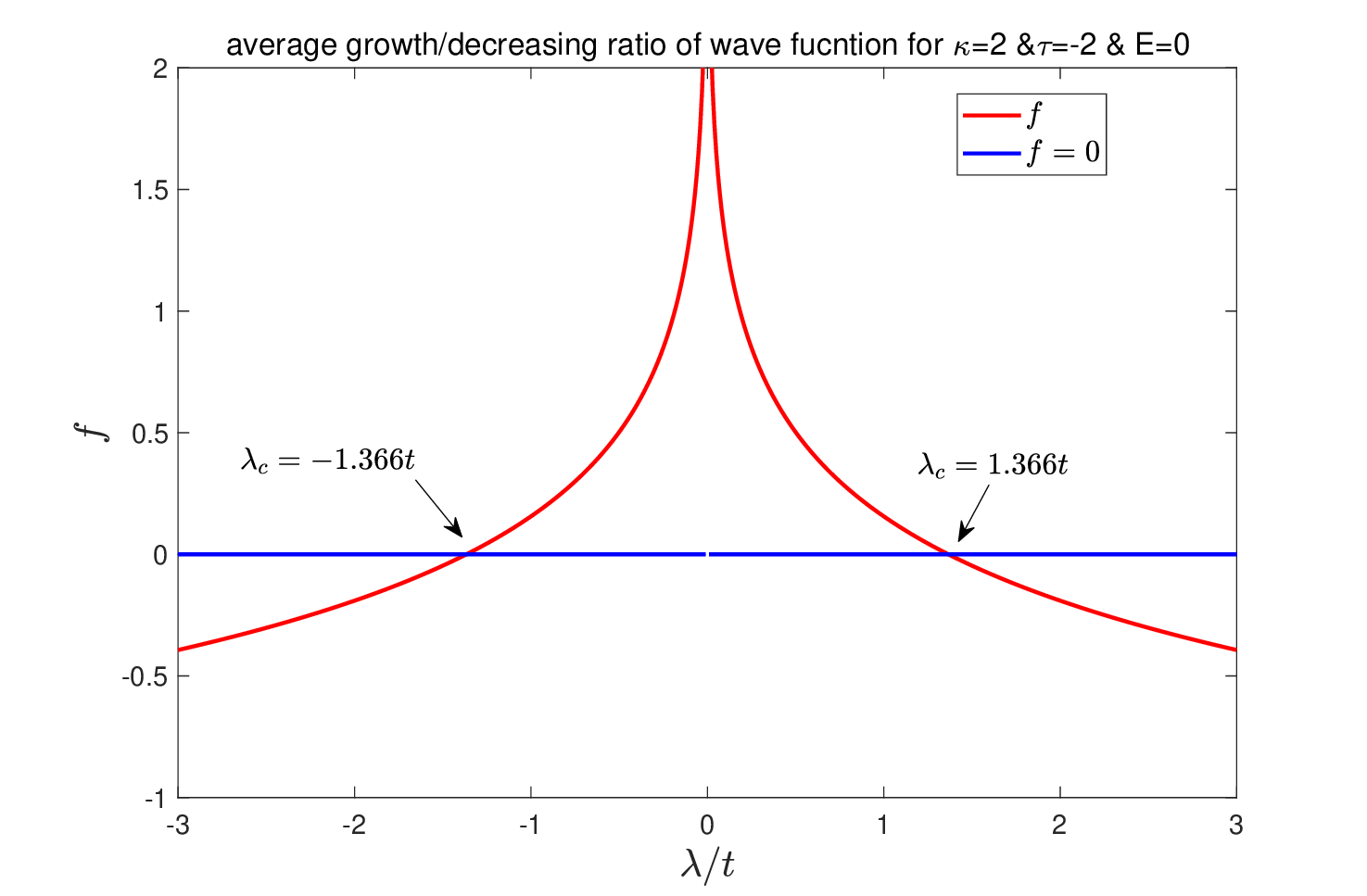}
\end{center}
\caption{ The average growth/decrease ratio of the zero-energy wave function for $\kappa=2$ and $\tau=-2$. The critical hopping strength is $\lambda=\lambda_c\simeq\pm1.366t$, where $f$ is exactly zero (indicated by black arrows in the figure).  }
\label{bu0}
\end{figure}

\begin{figure}
\begin{center}
\includegraphics[width=1.0\columnwidth]{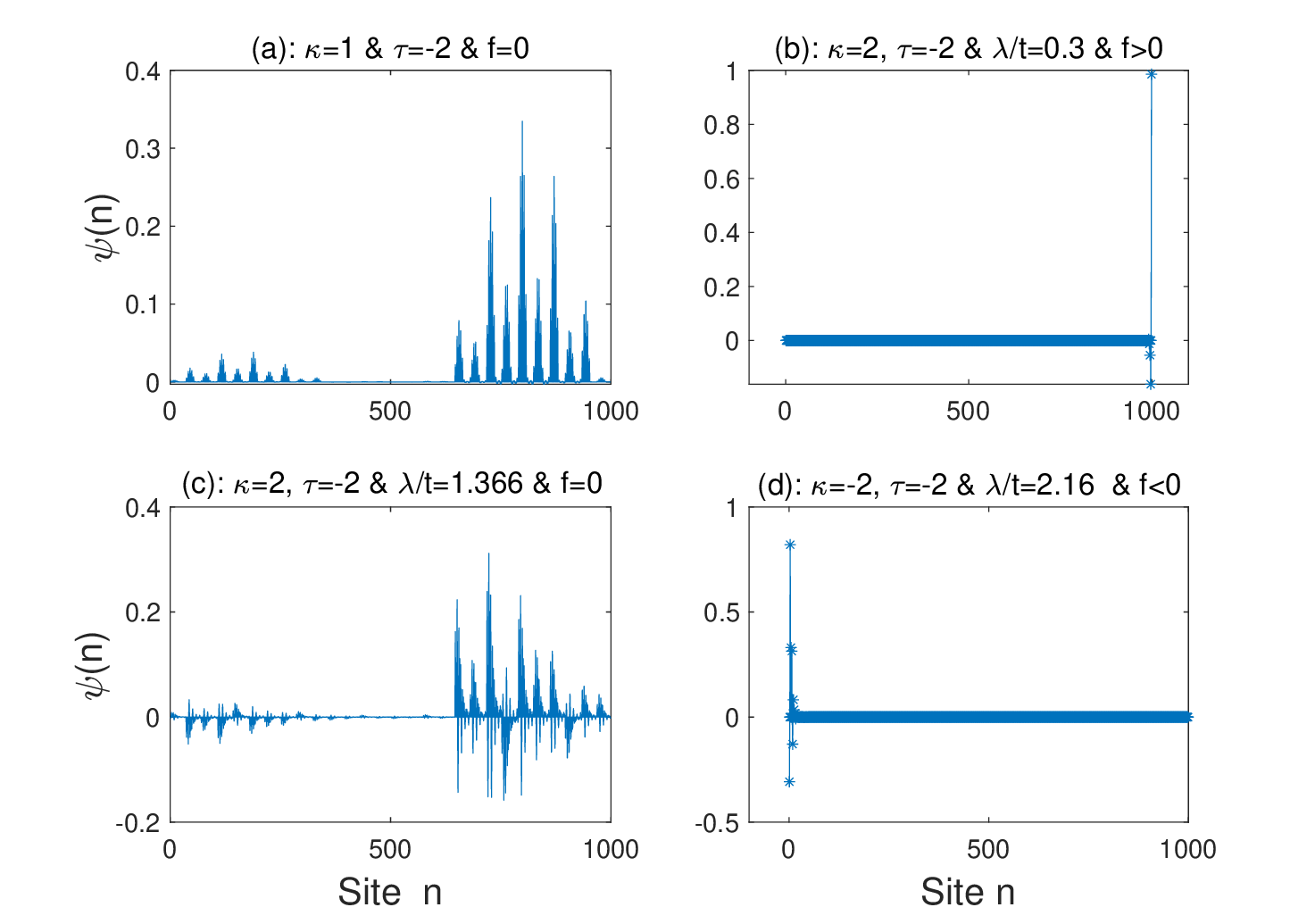}
\end{center}
\caption{Several typical zero-energy wave functions of critical states and localized edge states. Panels (a) and (c) show critical zero-energy wave functions where $f=0$. Panel (b) and (d) display localized right-hand and left-hand edge states where $f>0$ and $f<0$, respectively.}
\end{figure}

\section{localization of zero-energy state}
 In this section, we will discuss the influence of the parity of the integer $\kappa$ on the localization properties of zero-energy states. For the quasiperiodic model in Eq. (1), we observe that applying the transformation $\psi(n) \rightarrow (-1)^n \psi(n)$ would result in a change in energy sign, i.e., $E \rightarrow -E$. This is due to the chiral (sublattice) symmetry, where the energies $E_n$ and $-E_n$ appear in pairs \cite{Inui,Cheraghchi2005,Tabanelli}. Furthermore, the number of eigenenergies is equal to the number of lattice sites. If the total number of lattice sites $N$ is odd, then there will be at least one zero-energy state. However, we have found that when $N$ is even, there are usually no zero-energy eigenstates. Therefore, in this section, we assume that the total number of lattice sites $N$ is odd, ensuring the existence of a zero energy state.

Furthermore, we assume the lattice sites of system are labeled with number $i=1,2,3,...,2m, N=2m+1$, where $m$ is a positive integer.
Starting from wave functions of left-hand end site $\psi(i=1)=1$ [and $\psi(i=0)=0$], by Eq.(1),
the wave function of zero-energy state can be written as
\begin{align}
&\psi(N=2m+1)=\frac{V_{2m,2m-1}V_{2m-2,2m-3}...V_{4,3}V_{2,1}}{V_{2m,2m+1}V_{2m-2,2m-1}...V_{4,5}V_{2,3}}\psi(i=1),\notag\\
&=\frac{v(2m-1)v(2m-3)...v(3)v(1)}{v(2m)v(2m-2)...v(4)v(2)}.
\end{align}
In the above equation, we set $v(i) \equiv V_{i,i+1}$ and use the relation $V_{i+1,i}=V_{i,i+1}$.
The average growth/decrease ratio of the wave function can be expressed as follows:
\begin{align}
&f=\lim_{m\rightarrow\infty}\frac{1}{2m}ln(|\frac{\psi(N=2m+1)}{\psi(i=1)}|)\notag\\
&=\lim_{m\rightarrow\infty}\frac{1}{2m}ln(|\frac{v(2m-1)v(2m-3)...v(3)v(1)}{v(2m)v(2m-2)...v(4)v(2)}|).
\end{align}
If $f\geq0$,  then $f$ can be considered as the Lyapunov exponent $\gamma(E)$  (see Sec. \textbf{III}).

\subsection{$\kappa$ is a positive even integer}
  When  mosaic period $\kappa$ is a positive even integer, we can assume that $N = n\kappa + 1$, where $n$ is an integer. Using equations (1) and (5), we can see that the ergodicity of the map $\phi \longrightarrow 2\pi\beta i + \phi$ allows us to reduce the average growth / decrease ratio of the wave function to:
\begin{align}
&f=\lim_{n\rightarrow\infty}\frac{1}{n\kappa}ln(\frac{1}{|v(\kappa)v(2\kappa)...v(n\kappa)|}),\notag\\
&=\frac{-1}{\kappa\times 2\pi}[\int_{0}^{2\pi}d\phi ln(|v(\kappa,\phi)|)],\notag\\
&=\frac{-1}{\kappa}ln(\frac{2|\lambda/t|}{1+\sqrt{1-\tau}}),
\end{align}
where $v(\kappa,\phi)\equiv \frac{2\lambda \cos(2\pi\beta\kappa+\phi)}{\sqrt{1-\tau \cos^2(2\pi\beta\kappa+\phi)}}$ [see Eq.(2)].

It is shown that when $\kappa$ is a positive even integer, for a generic  $\lambda$, $f$ is usually not zero. Then the zero-energy state would be localized states which may situate at right-hand edge ($f>0$) or left-hand edge ($f<0$) of the lattices (see Figs.1 and 2). So for general parameters, the zero-energy state would be a localized edge state.
Only  when $f$ is exactly vanishing, i.e., $f=0$,
\begin{align}
&\rightarrow f=\frac{-1}{\kappa}ln(\frac{2|\lambda/t|}{1+\sqrt{1-\tau}})=0\notag\\
&\rightarrow |\lambda/t|=|\lambda_c/t|\equiv\frac{1+\sqrt{1-\tau}}{2},
\end{align}
 the zero-energy state would be  a critical state as shown in Figs. 1 and 2. Fig. 1 displays the average growth/decrease ratio, denoted as $f$, for $\kappa=2$ and $\tau=-2$. At the critical strength of $\lambda_c\simeq\pm 1.366t$, the value of $f$ is equal to 0. As $\lambda$ approaches the critical value of $\lambda_c$, the localization length can become infinitely large, represented by $\xi\equiv 1/|f|\propto 1/|\lambda-\lambda_c|\rightarrow\infty$.

 To investigate the properties of zero-energy states, we numerically solve Eq. (1) for $\kappa=2$, $\tau=-2$, and a lattice size of $N=2\times500+1$. In Fig. 2, we present several typical wave functions for localized and critical states. It is known that the wave function of an extended state typically spans the entire lattice, while a localized state only occupies a finite number of lattice sites. The critical state is composed of several disconnected patches that interpolates between the localized and extended states \cite{yicai,Liu2022,zhangyicai2024}.
As shown in Fig. 2, the zero-energy wave function for $f=0$ is a critical state. The wave functions with non-zero values of $f$ correspond to localized edge states. Specifically, when $f>0$, the state is located at the right-hand end edge, and when $f<0$, it is located at the left-hand end edge.

\subsection{$\kappa$ is a positive odd integer}
When $\kappa$ is a positive odd integer, we can assume $N=2n\kappa+1$, where $n$ is an integer.
  Similarly, $f$ can be written as
\begin{align}
&f=\lim_{n\rightarrow\infty}\frac{1}{2n\kappa}ln(\frac{|v(\kappa)v(3\kappa)...v((2n-1)\kappa)|}{|v(2\kappa)v(4\kappa)...v(2n\kappa)|}),\notag\\
&=\frac{1}{2\kappa\times 2\pi}[\int_{0}^{2\pi}d\phi ln(|v(\kappa,\phi)|)-\int_{0}^{2\pi}d\phi ln(|v(2\kappa,\phi)|)],\notag\\
&=0.
\end{align}
It shows that when $\kappa$ is a positive odd integer, the average growth/decrease ratio of the zero-energy wave function is exactly zero. This means that if mosaic period $\kappa$ is odd, all of the zero-energy states are critical states.

It is known that there exists a Thouless's relation between Lyapunov exponent and energy spectrum in one-dimensional lattice model only with nearest neighbor hopping \cite{Thouless1972}, i.e.,
\begin{align}\label{9}
&\gamma(E_n)\notag\\
&=\lim_{N\rightarrow \infty}\{\frac{1}{N-1}[\sum_{n'\neq n}ln|E_n-E_n'|-ln|V_{1,2}V_{2,3}...V_{N-1,N}|]\},
\end{align}
where $E_n$ is the $n-th$ eigenvalue of Hamiltonian and $N$ is lattice size.
Now we consider zero-energy states, then the average grown ratio $f$ in Eqs.(6) and (8) can be identified with Lyapunov exponent $\gamma(E_n=0)$ here. Then we get
\begin{align}\label{81}
f=\lim_{N\rightarrow \infty}\{\frac{1}{N-1}[\sum_{E_n\neq 0}ln|E_n|-ln|V_{1,2}V_{2,3}...V_{N-1,N}|]\}.
\end{align}

Now there are two cases for even and odd $\kappa$.

\textbf{Case I: $\kappa$ is an even integer}

Comparing with Eq.(6), we can see that the second term of the right-hand sides in Eq.(\ref{81}) is exactly $f$ itself. This leads to a sum rule for the energy spectrum, as follows:
\begin{align}\label{82}
\lim_{N\rightarrow \infty}\frac{1}{N-1}\sum_{E_n\neq 0}ln|E_n|=0.
\end{align}
 Because we use unit system of the constant hopping $t=1$ [see Eq.(2)], then Eq.(\ref{82}) implies that, when $\kappa$ is even integer, the geometric mean value of the absolute values of energy spectrum is equal to the constant hopping $t$.

\textbf{Case II: $\kappa$ is an odd integer}

From Eq.(8), we know $f=0$ , then by Eq.(\ref{81}), the sum rule for odd $\kappa$ is reduced to
\begin{align}\label{83}
&\lim_{N\rightarrow \infty}\frac{1}{N-1}\sum_{E_n\neq 0}ln|E_n|\notag\\
&=\lim_{N\rightarrow \infty}\frac{1}{N-1}ln|V_{1,2}V_{2,3}...V_{N-1,N}|.
\end{align}
This shows that when $\kappa$ is an odd integer, the geometric mean value of the absolute values of energy spectrum is equal to the geometric mean value of the hopping.

Similar to Eq. (6), when $N$ is very large, we can use an integration to evaluate the right-hand side of the above Eq.(\ref{83}). 
Therefore, the sum-rule of energy spectrum for odd $\kappa$ is reduced to
\begin{align}\label{84}
&\lim_{N\rightarrow \infty}\frac{1}{N-1}\sum_{E_n\neq 0}ln|E_n|\notag\\
&=\frac{1}{\kappa\times2\pi}[\int_{0}^{2\pi}d\phi ln(|v(\kappa,\phi)|)]=\frac{1}{\kappa}ln(\frac{2|\lambda/t|}{1+\sqrt{1-\tau}}).
\end{align}
The above two sum-rules of energy spectrum for even and odd $\kappa$, i.e., Eqs.(\ref{82}) and (\ref{84})  are verified by our numerical calculations.

Some interesting odd-even effects of the lattice site number $N$ have been investigated in random off-diagonal disorder models. It has been discovered that a delocalization transition only occurs when the lattice size $N$ is odd \cite{Brouwer1998}. Additionally, the localization length of the zero-energy state is highly dependent on the boundary conditions \cite{Brouwer2002}, and it can reach arbitrarily large values.
Furthermore, density of state near zero energy shows a singularity that strongly depends on the parity of lattice size $N$ \cite{Brouwer2000}.


 The above discussions show that the parity of integer $\kappa$ has important influences on the localization of zero-energy states.
  An odd integer $\kappa$ leads to a critical zero-energy state, whereas an even $\kappa$ typically results in localized edge states (refer to Fig. 2). It should be noted that the aforementioned discussion on the odd-even effects and energy spectrum's sum-rules can also be extended to other forms of quasiperiodic hopping with mosaic modulations \cite{Ganeshan2013, Zeng2021} .

In the following text, we will show that the above influences of parity of $\kappa$ are also  extend to other eigenstates in the vicinity of zero energy.
Specifically, when $\kappa$ is a positive odd integer and the hopping strength $\lambda$ is fixed, the eigenstates near zero energy are always critical (see Sec. \textbf{IV}). This means that if the energy is sufficiently close to zero, there will be no Anderson localization transition for odd $\kappa$. On the other hand, when $\kappa$ is a positive even integer, the eigenstates near zero energy will undergo an Anderson localization transition as the quasiperiodic hopping strength increases. As a result, the system will have localized states near zero energy.

\section{The  Lyapunov exponent  }
  When $E\neq0$, the localization properties of eigenstates can be characterized by the Lyapunov exponent. In this section, we will use the transfer matrix method \cite{Sorets1991,Davids1995} to calculate the Lyapunov exponent.

First of all,   let us assume that the system is a half-infinite lattice with left-hand end sites at $i=0$ and $i=1$.
Further using Eq.(1), starting with $\psi(0)$ and $\psi(1)$ of left-hand end sites, we can  calculate the wave function for the entire system with relation
\begin{align}
\Psi(i)=T(i)T(i-1)...T(2)T(1)\Psi(0)
\end{align}
where transfer matrix
\begin{align}\label{V}
T(n)\equiv\left[\begin{array}{ccc}
\frac{E}{V_{n,n+1}} &-\frac{V_{n,n-1}}{V_{n,n+1}}  \\
1&0\\
  \end{array}\right].
\end{align}
and
\begin{align}
\Psi(n)\equiv\left[\begin{array}{ccc}
\psi(n+1)  \\
\psi(n)\\
  \end{array}\right].
\end{align}

For a given parameter $E$, as $n$ increases, it can be assumed that the wave function grows exponentially according to a law \cite{Ishii,Furstenberg}, expressed as
\begin{align}
\psi(n)\sim e^{\gamma(E) n}, &\ as \ n\rightarrow \infty,
\end{align}
where $\gamma(E)\geq0$ is Lyapunov exponent which measures the average growth rate of wave function.
If the parameter $E$ is not an eigen-energy of $H$, the Lyapunov exponent will be positive, i.e., $\gamma(E)>0$ \cite{Jonhnson1986}. When $E$ is an eigen-energy of the system, the Lyapunov exponent can be either zero or positive \cite{yicai}. For critical states, the Lyapunov exponent is equal to zero, while for localized states, the Lyapunov exponent is greater than zero.

Consequently the Lyapunov exponent can be written as
\begin{align}
&\gamma(E)=\lim_{L \rightarrow \infty }\frac{\log(|\Psi(L)|/|\Psi(0)|)}{L}\notag\\
&=\lim_{L\rightarrow \infty}\frac{\log(|T(L)T(L-1)...T(2)T(1)\Psi(0)|/|\Psi(0)|)}{L}
\end{align}
where $L$ is a positive integer and
\begin{align}
|\Psi(n)|=\sqrt{|\psi(n+1)|^2+|\psi(n)|^2}.
\end{align}

In the following, we consider the adjacent $\kappa$ lattice sites as a ``super unit cell".  Additionally, we assume that $L=m\kappa+1$ (where $m$ is an integer) and that $|\Psi(0)|/|\Psi(1)|$ is a finite, non-zero real number. In this case, the Lyapunov exponent can be reduced to
\begin{align}
&\gamma(E)=\lim_{L \rightarrow \infty }\frac{\log(|\Psi(L)|/|\Psi(0)|)}{m\kappa+1}\notag\\
&=\lim_{m\rightarrow \infty}\frac{\log(|T(m\kappa+1)T(m\kappa)...T(2)\Psi(0)|/|\Psi(0)|)}{m \kappa}\notag\\
&=\frac{1}{\kappa}\lim_{m\rightarrow \infty}\frac{\log(|CT_m.CT(m-1)...CT_1\Psi(0)|/|\Psi(0)|)}{m}\notag\\
\end{align}
where cluster transfer matrix $CT_n$  for $n-th$ ``super unit cell"  is defined  as
\begin{align}
&CT_n\equiv T(n\kappa+1)T(n\kappa)...T((n-1)\kappa+3)T((n-1)\kappa+2)\notag\\
&=\left[\begin{array}{ccc}
E &-v(n\kappa) \\
1&0\\
  \end{array}\right]\left[\begin{array}{ccc}
\frac{E}{v(n\kappa)} &-\frac{1}{v(n\kappa)} \\
1&0\\
  \end{array}\right]\left[\begin{array}{ccc}
E &-1 \\
1&0\\
  \end{array}\right]^{\kappa-2}\notag\\
  &=\frac{\left[\begin{array}{ccc}
E^2-(E^2\tau+4\lambda^2)cos^2(\theta_n) &-E(1-\tau cos^2(\theta_n) ) \\
E(1-\tau cos^2(\theta_n) )&-1+\tau cos^2(\theta_n) \\
  \end{array}\right]}{2\lambda cos(\theta_n)\sqrt{1-\tau cos^2(\theta_n)}}\notag\\
  &\times \left[\begin{array}{ccc}
E &-1 \\
1&0\\
  \end{array}\right]^{\kappa-2}
\end{align}
where $\theta_n=2\pi\beta n\kappa +\phi$.

The cluster transfer matrix Eq.(16) can be further
written as a product of two parts, i.e., $CT_n= A_nB_n$, where
   \begin{align}
&A_n=\frac{1}{2\lambda cos(\theta_n)\sqrt{1-\tau cos^2(\theta_n)}},\notag\\
&B_n=\left[\begin{array}{ccc}
B_{11} &B_{12} \\
B_{21}&B_{22}\\
  \end{array}\right]\left[\begin{array}{ccc}
E &-1 \\
1&0\\
  \end{array}\right]^{\kappa-2},
\end{align}
with $B_{11}=E^2-(E^2\tau+4\lambda^2)cos^2(\theta_n)$, $B_{21}=-B_{12}=E(1-\tau cos^2(\theta_n))$ and $B_{22}=-1+\tau cos^2(\theta_n)$.
Now  the Lyapunov exponent is
\begin{align}
\gamma(E)=\frac{1}{\kappa}[\gamma_A(E)+\gamma_B(E)],
\end{align}
where
\begin{align}
&\gamma_A(E)=\lim_{m\rightarrow \infty}\frac{\log(|A(m)A(m-1)...A(2)A(1)|)}{m}.
\end{align}
and $\gamma_B(E)$ are given by
\begin{align}
&\gamma_B(E)=\lim_{m\rightarrow \infty}\frac{\log(|B(m)B(m-1)...B(2)B(1)\Psi(0)|/|\Psi(0)|)}{m}.
\end{align}

In the following, we will utilize Avila's global theory \cite{Avila2015} to obtain the Lyapunov exponent and Avila's acceleration (see next section). As suggested by Refs. \cite{Liu2021, YONGJIAN1}, our first step is to complexify phase $\phi\rightarrow \phi+i \epsilon$ with $\epsilon >0$ , e.g., $A_n=\frac{1}{2\lambda cos(2\pi \beta n\kappa +\phi+i\epsilon)\sqrt{1-\tau cos^2(2\pi\beta n\kappa +\phi+i\epsilon)}}$.
 In addition, due to the ergodicity of the map $\phi\longrightarrow 2\pi\beta n+\phi$, we can write  $\gamma_A(E)$ as an integral over phase $\phi$ \cite{Longhi2019}, consequently
\begin{align}
&\gamma_A(E,\epsilon)\notag\\
&=\frac{1}{2\pi}\int_{0}^{2\pi} d\phi \ln(|\frac{1}{2\lambda cos(\phi+i\epsilon)\sqrt{1-\tau cos^2(\phi+i\epsilon)}}|)\notag\\
&=-\epsilon+\ln(|\frac{2}{\lambda(1+\sqrt{1-\tau})}|),
\end{align}
for $\epsilon< \ln |\frac{2+2\sqrt{1-\tau}-\tau}{\tau}|$.

Next we take $\epsilon\rightarrow\infty$
 \begin{align}
&B_n=\frac{e^{-i(4\pi\beta n \kappa+\phi)+2\epsilon}}{4}\left[\begin{array}{ccc}
-( E^2 \tau+4\lambda^2) &E\tau \\
-E\tau & \tau\\
  \end{array}\right]\left[\begin{array}{ccc}
E & -1 \\
1& 0\\
  \end{array}\right]^{\kappa-2}\notag\\
  &+O(1).
\end{align}
Then for large $\epsilon$, i.e.,  $\epsilon\gg1$, $\gamma_B(E,\epsilon)$ is determined by the largest eigenvalue (in absolute value) of $B_n$, i.e.,
\begin{align}
\gamma_B(E,\epsilon)=2\epsilon+\ln(|\frac{|P|+\sqrt{P^2+16\lambda^2\tau}}{8}|),
\end{align}
where
\begin{align}
P=(\tau E^2+4\lambda^2)a_{\kappa}+\tau a_{\kappa-2}-2\tau E a_{\kappa-1}.
\end{align}
and $a_{\kappa}$, is given by
\begin{align}
a_\kappa=\frac{1}{\sqrt{E^2-4}}[(\frac{E+\sqrt{E^2-4}}{2})^{\kappa-1}-(\frac{E-\sqrt{E^2-4}}{2})^{\kappa-1}].
\end{align}

\begin{figure}
\begin{center}
\includegraphics[width=1.0\columnwidth]{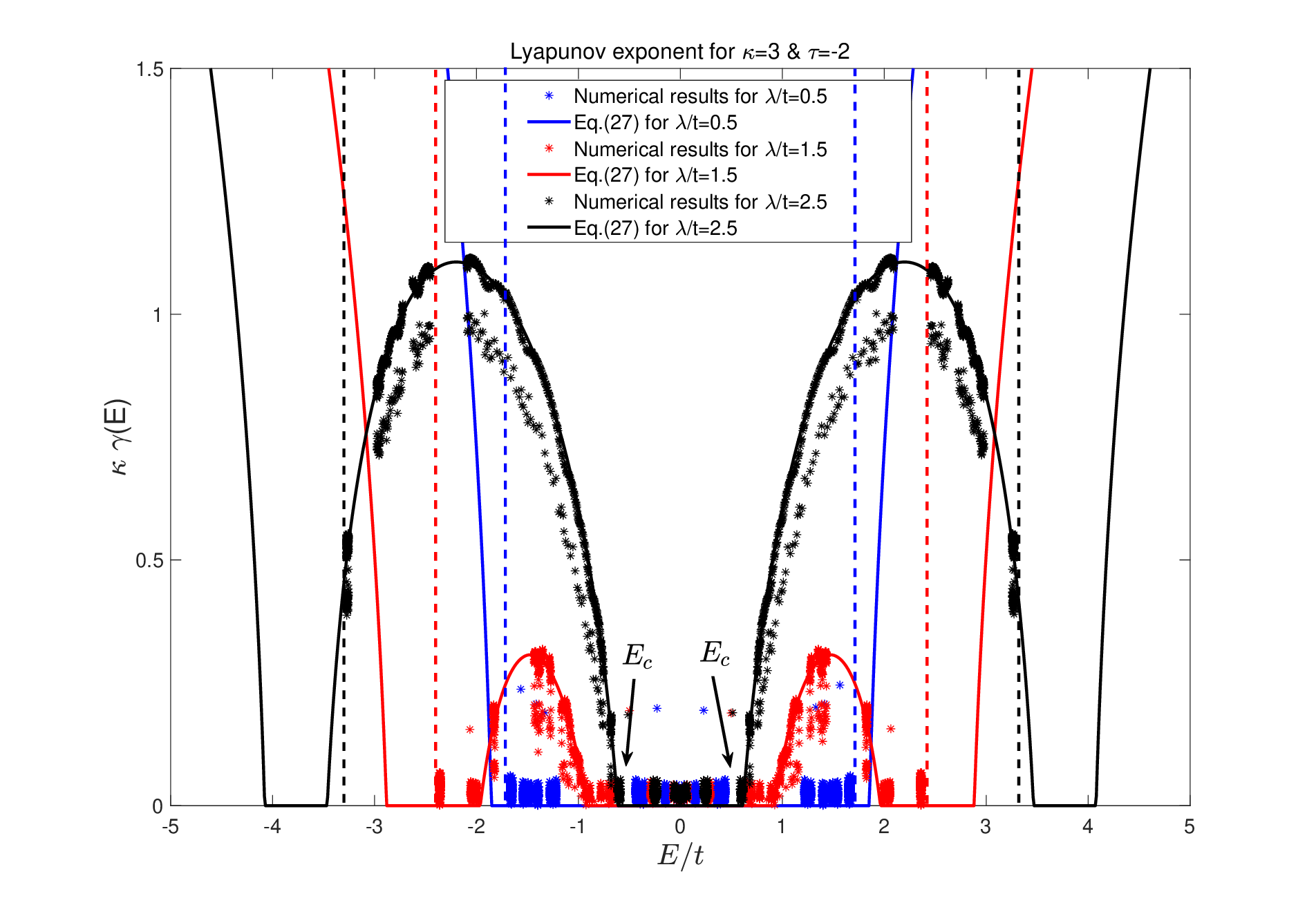}
\end{center}
\caption{ Lyapunov exponents are calculated for $\kappa=3$, $\tau=-2$, and $\lambda/t=0.5, 1.5, 2.5$. The numerical results for all the eigenenergies are represented by discrete points, while the solid lines correspond to Eq.(27). The mobility edges for $\lambda/t=2.5$ are denoted by black arrows. As the energy approaches the mobility edges of the localized-critical transition, the Lyapunov exponent $\gamma(E)$ tends to zero as $E$ approaches $E_c$. The critical index of the localization length is $\nu=1$. We also label the energy band edges with dashed lines in this figure.  }
\end{figure}

\begin{figure}
\begin{center}
\includegraphics[width=1.0\columnwidth]{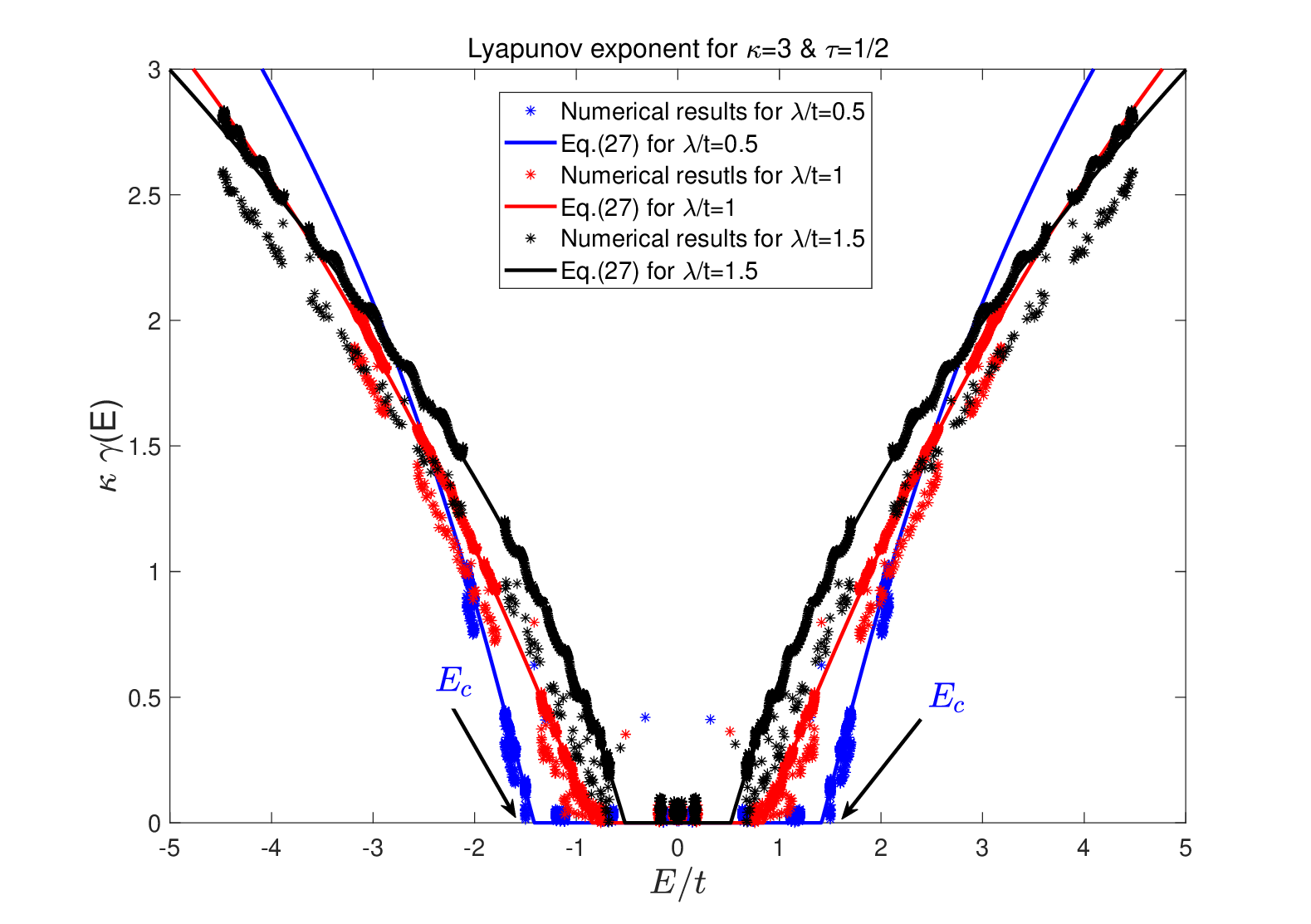}
\end{center}
\caption{ For $\kappa=3$ and $\tau=1/2$, we calculated Lyapunov exponents for different values of $\lambda/t$, specifically for $\lambda/t=0.5, 1.0, 1.5$. The numerical results for all the eigenenergies are represented by discrete points, while the solid lines correspond to Eq.(27). In the case of $\lambda/t=0.5$, the mobility edges are denoted by black arrows. Near the mobility edges of the localized-critical transition (e.g., $E_c/t\simeq\pm \sqrt{2}$ for $\lambda/t=0.5$), the Lyapunov exponent $\gamma(E)\propto |E-E_c|$ approaches zero as $E$ approaches $E_c$. Moreover, the critical index of the localization length is $\nu=1$.}
\end{figure}
When $\epsilon$ is very small, using the facts that $\gamma(E,\epsilon)\geq0$  and $\gamma_B(E,\epsilon)$ is  a convex and piecewise linear function of $\epsilon$ \cite{Avila2015,YONGJIAN1},  one can get
\begin{align}
&\gamma(E,\epsilon)=Max\{0,\gamma_A(E,\epsilon)+\gamma_B(E,\epsilon)\},\notag\\
&=
\frac{1}{\kappa}Max\{0,\epsilon+\ln(|\frac{|P|+\sqrt{P^2+16\lambda^2\tau}}{4\lambda(1+\sqrt{1-\tau})}|)\}
\end{align}
Furthermore, when $\epsilon=0$, the Lyapunov exponent $\gamma(E)\equiv\gamma(E,\epsilon=0) $ is
\begin{align}
\gamma(E)=\frac{1}{\kappa}Max\{\ln|\frac{|P(E)|+\sqrt{P^2(E)+16\lambda^2\tau}}{4\lambda(1+\sqrt{1-\tau})}|,0  \}
\end{align}
where
\begin{align}
P(E)=(\tau E^2+4\lambda^2)a_{\kappa}+\tau a_{\kappa-2}-2\tau E a_{\kappa-1}
\end{align}
and
\begin{align}
a_\kappa=\frac{1}{\sqrt{E^2-4}}[(\frac{E+\sqrt{E^2-4}}{2})^{\kappa-1}-(\frac{E-\sqrt{E^2-4}}{2})^{\kappa-1}].
\end{align}
and integer $\kappa\geq2$.

When $\tau=0$, then $P(E)=4\lambda^2a_{\kappa}$, and
\begin{align}
\gamma(E)=\frac{1}{\kappa}Max\{\ln|\lambda a_{\kappa}|,0  \}.
\end{align}

The formula Eq. (27) has been confirmed by our numerical results, as shown in Figs. 3 and 4.
In order to accurately calculate the Lyapunov exponents, two conditions must be met: firstly, the integer $L$ must be sufficiently large; and secondly, in order to guarantee most of discrete points falling on the solid lines [the exact results given by Eq.(27)],    $L$ should also be significantly smaller than the system size $N$, specifically $1 \ll L \ll N$.

 To be specific, when we set $\kappa=3$ and $\tau=-2,1/2$, with a system size of $N=3\times1000$, we obtain the corresponding eigenenergies and eigenstates. We then proceed to numerically calculate the Lyapunov exponents for all of these eigenenergies, as shown in the discrete points in Figs. 3 and 4.
 In our numerical calculation, we take $L=200$,  phase $\phi=0$, $\psi(0)=0$ and $\psi(1)=1$ in Eq.(13).
   The solid lines in Figs. 3 and 4 are obtained using the same parameters in Eq. (27). It is evident that the majority of the discrete points fall onto the solid lines.

 However, we also observe that there are certain localized states that do not fall on the solid lines. This is due to the fact that these wave functions are located too close to the left-hand boundary of the system.

\section{mobility edge and critical region }
 The mobility edges, denoted by $E_c$, separate the localized states from the critical states and can be determined using Eq. (27). The condition for determining $E_c$ is as follows:
\begin{align}
\gamma(E=E_c)=\frac{1}{\kappa}\ln|\frac{|P(E)|+\sqrt{P^2(E)+16\lambda^2\tau}}{4\lambda(1+\sqrt{1-\tau})}|=0
\end{align}
then
\begin{align}
|P(E=E_c)|=4|\lambda|\sqrt{1-\tau},
\end{align}
The critical regions, which consist of critical states, are determined by
\begin{align}
|P(E)|<4|\lambda|\sqrt{1-\tau}.
\end{align}

\begin{figure}
\begin{center}
\includegraphics[width=1.0\columnwidth]{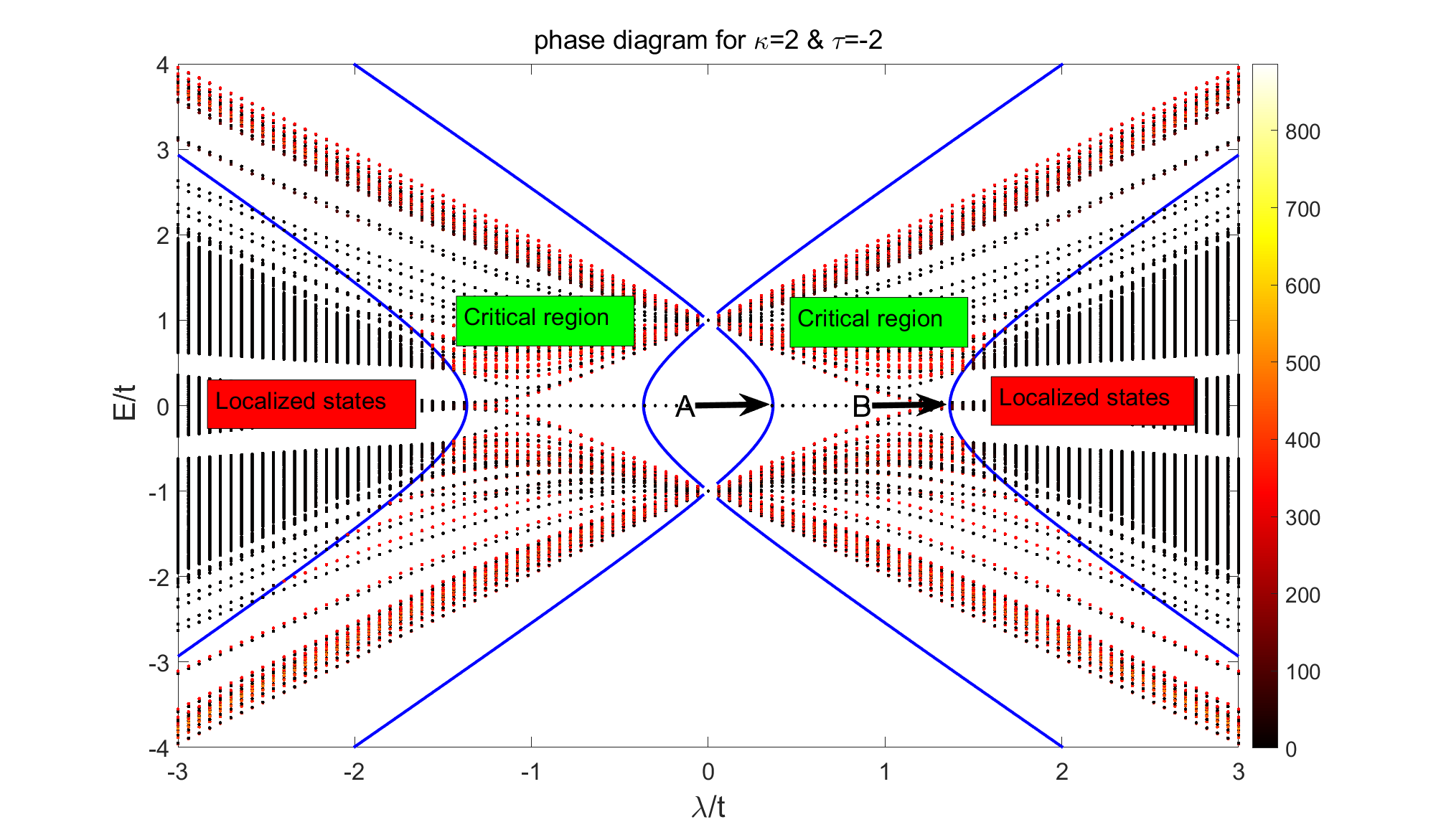}
\end{center}
\caption{ In the $(\lambda, E)$ plane, the phase diagram for $\kappa=2$ and $\tau=-2$ shows that when $E$ is close to zero, localized-critical transitions exist. The phase boundaries (mobility edges $E_c$) are indicated by the blue solid lines, and their equations are given by Eq. (32).
 Standard deviations are represented with different colors.
Note: here, we set the total lattice site number  N to be 500*2-1=999, an odd integer. As a result, zero energy localized states emerge [for the reasons please see the explanations of Sec.\textbf{II}].   }
\end{figure}

By expanding the Lyapunov exponent near the mobility edges $E_c$, we get
\begin{align}
\gamma(E)\propto |E-E_c|\rightarrow0, \ as \ E \rightarrow E_c.
\end{align}
Then the localization length is
\begin{align}
\xi(E)\equiv1/\gamma(E)\propto |E-E_c|^{-1}\rightarrow\infty,\ as \ E \rightarrow E_c.
\end{align}
Its critical index is 1, as shown by the finite slopes of the solid lines near $E_c$ in Figs. 3 and 4.

In order to further distinguish the localized states from the  critical states, we also numerically calculate standard deviation of coordinates of eigenstates \cite{Boers2007}
\begin{align}\label{37}
&\sigma=\sqrt{\sum_{i}(i-\bar{i})^2|\psi(i)|^2},
\end{align}
where the average value of coordinate $\bar{i}$ is
\begin{align}
\bar{i}=\sum_{i}i|\psi(i)|^2.
\end{align}
The standard deviation $\sigma$ describes the spatial extension of the wave function in the lattice. The phase diagram in the $[\lambda-E]$ plane is shown in Figs.  5,  6, and 7.   In Figs. 5,  6 and 7, the standard deviations of coordinates are represented with different colors.
It is evident from these figures that for localized states, the standard deviations of coordinates are significantly small, while for critical states, the standard deviations are considerably larger.

From Figs. 5, 6, and 7, it is evident that when $\kappa\geq 2$, there are $\kappa-1$ loops in the phase diagram for small hopping strength $\lambda/t$. Within these loops, the Lyapunov exponent is positive, indicating that $\gamma(E)>0$. This suggests that any eigenstates within the loops would be localized. However,  numerical results show that there are no eigenenergies falling within the loops (except for zero energy states when $N$ is odd integer[see Fig.5]).

When $\lambda/t=0$, the system exhibits $\kappa$ eigenenergies with multiple degeneracies. This is due to the fact that, at this point, the system is composed of identical independent unit cells, each containing $\kappa$ lattice sites. Within each unit cell, there are exactly $\kappa$ eigenenergies. As $\lambda/t$ increases, the degeneracies are lifted and the system enters the critical region. Furthermore, the parity of $\kappa$ also plays a significant role in the phase diagram.

\subsection{$\kappa$ is an even number}
When $\kappa$ is an even number and the energy $E$ is very close to zero, Anderson localizations occur if the potential hopping strength $\lambda$ is sufficiently large (see Fig. 5 for $\kappa=2$). In particular, as $E$ approaches zero, from Eq.(32), we get two critical hopping strengths:
\begin{align}
\lambda_{c1}=\pm \frac{1-\sqrt{1-\tau}}{2} \ \ \& \ \ \lambda_{c2}=\pm \frac{1+\sqrt{1-\tau}}{2}.
\end{align}
Here, the values of $\lambda_{c1}$ and $\lambda_{c2}$ are independent of $\kappa$. In Fig. 5, $\lambda_{c1}$ corresponds to point $A$ and $\lambda_{c2}$ corresponds to point $B$. It is worth noting that $\lambda_{c2}$ is the same as the critical value $\lambda_c$ mentioned in Section II, where the average growth rate of the zero-energy wave function is zero [see Eq.(7)].

From Fig. 5, it is evident that when the hopping parameter $\lambda$ falls within the range of $|\lambda_{c1}|<\lambda<|\lambda_{c2}|$, the eigenstates are critical states. However, when $\lambda$ falls outside of this interval, the eigenstates near zero energy become localized. This indicates  the presence of Anderson localization transitions when $\kappa$ is an even number.

%
%

\begin{figure}
\begin{center}
\includegraphics[width=1.0\columnwidth]{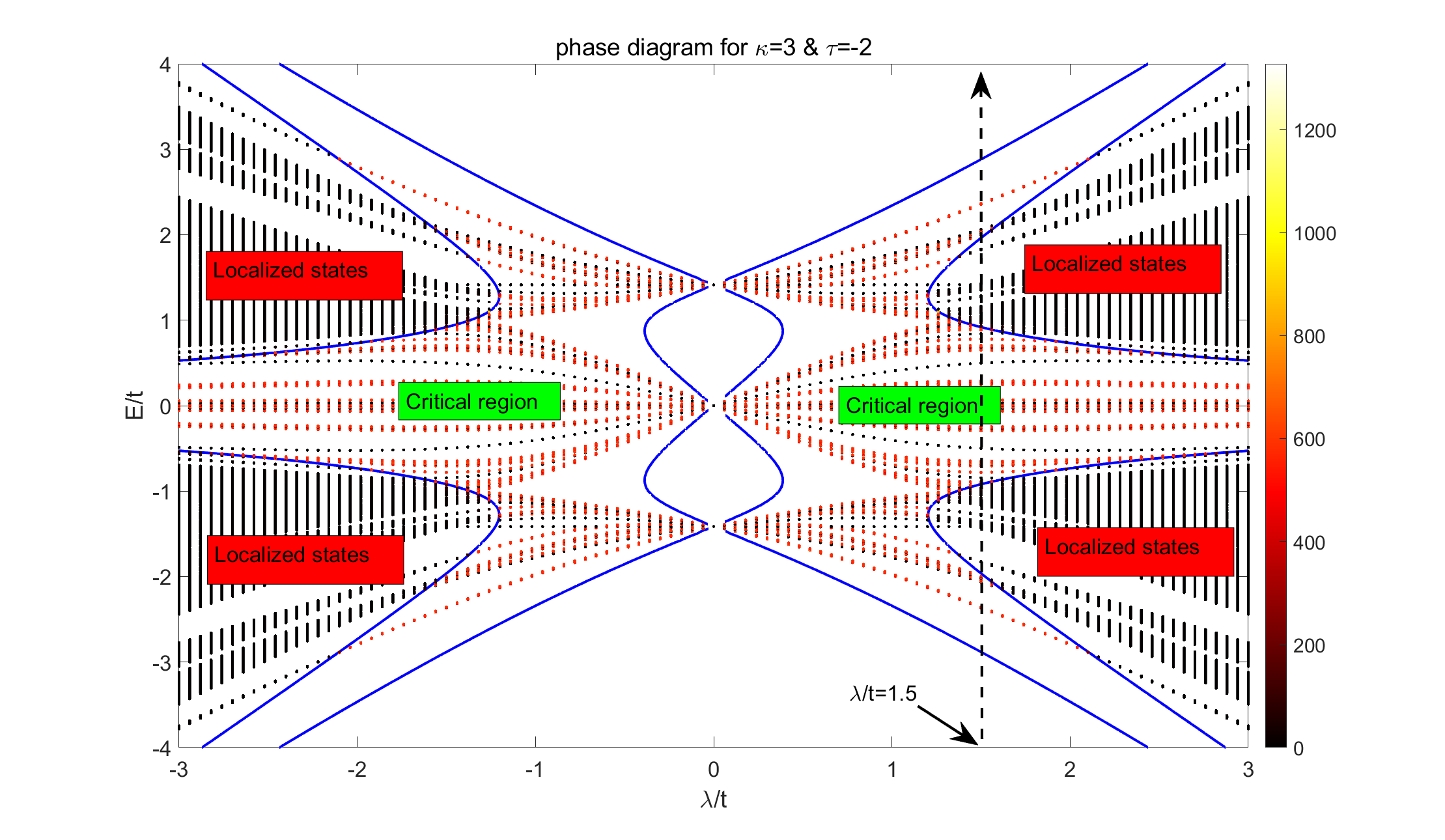}
\end{center}
\caption{ Phase diagram in the $(\lambda, E)$ plane for $\kappa=3$ and $\tau=-2$. When $E$ is near zero, there are no localized-critical transitions. The blue solid lines represent the phase boundaries (mobility edges $E_c$), which are determined by Eq. (32).
 Standard deviations are represented with different colors. Here we take lattice site number $N=1000*3=3000$.  }
\end{figure}
\begin{figure}
\begin{center}
\includegraphics[width=1.0\columnwidth]{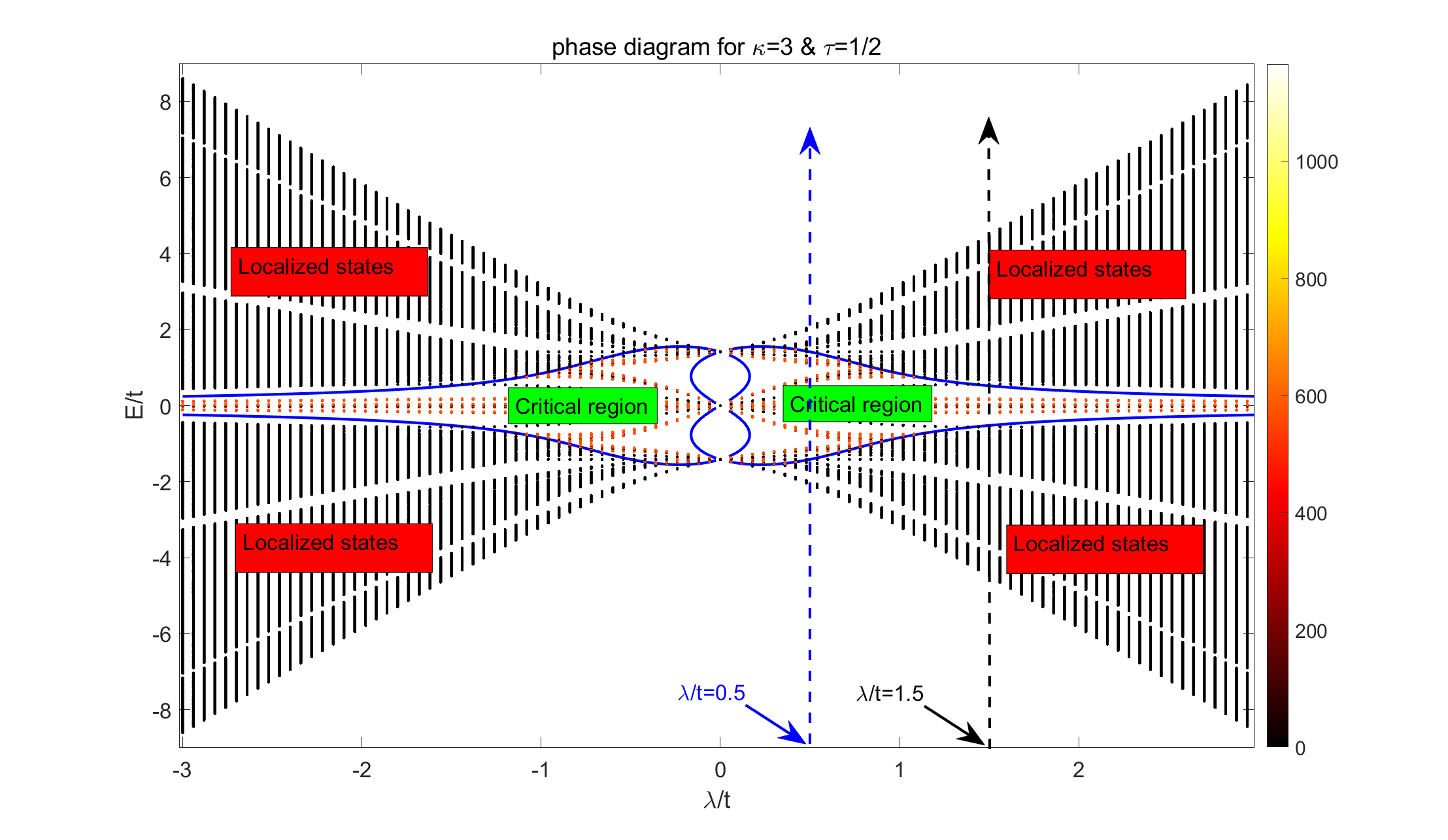}
\end{center}
\caption{ Phase diagram in the $(\lambda, E)$ plane for $\kappa=3$ and $\tau=1/2$. There are localized-critical transitions for nonzero energy states. The phase boundaries (mobility edges $E_c$) are represented by blue solid lines, given by Eq.(32).
 Standard deviations are represented with different colors. Here we take lattice site number $N=1000*3=3000$.}
\end{figure}

\subsection{$\kappa$ is an odd number}
%

\begin{figure}
\begin{center}
\includegraphics[width=1.0\columnwidth]{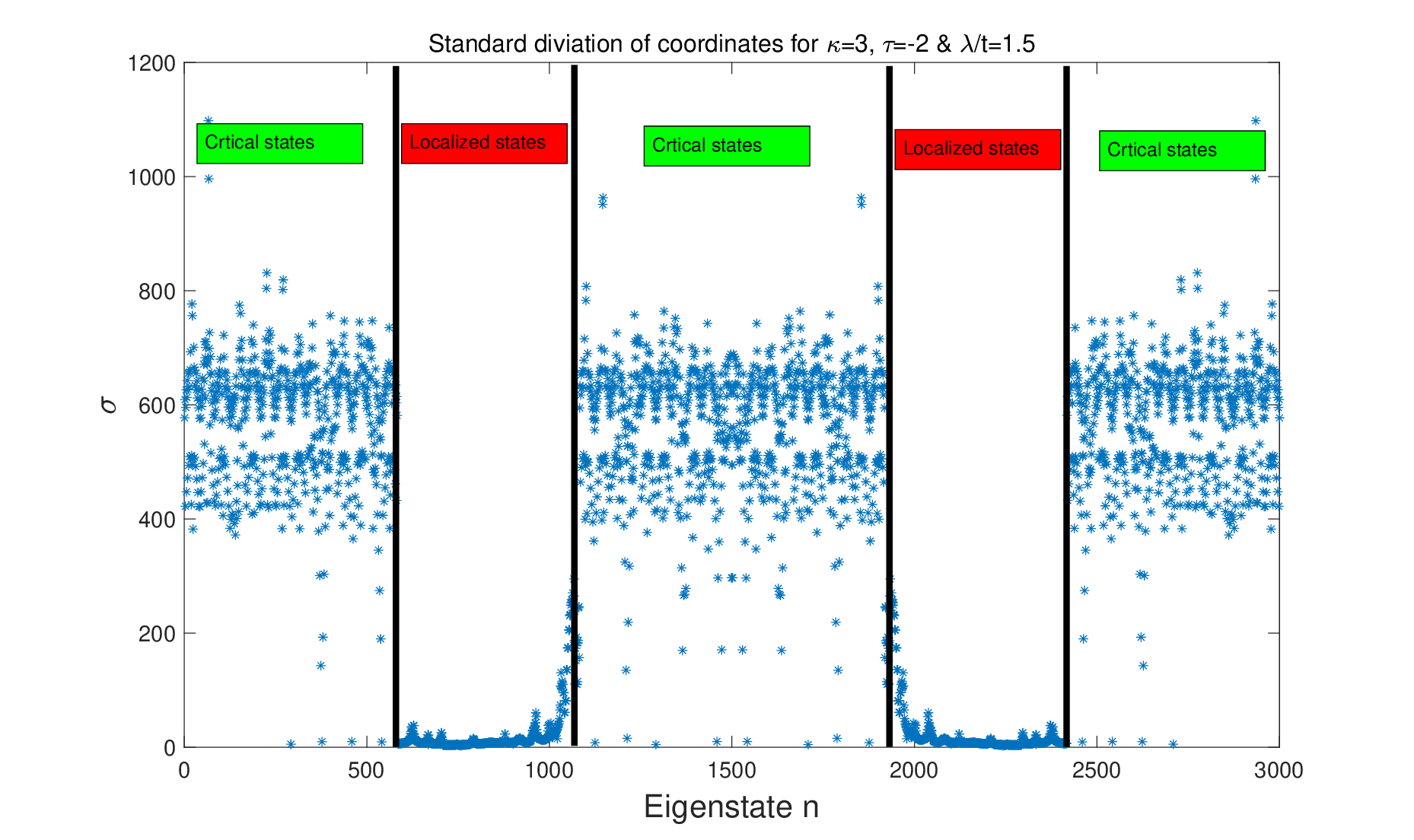}
\end{center}
\caption{ Standard deviations of localized states and critical states for parameters $\kappa=3$, $\tau=-2$, and $\lambda/t=1.5$. The eigenenergy $E_n$ gradually increases as the eigenstate index $n$ ranges from $1$ to $3000$, following the black dashed line in Fig. 6.}
\end{figure}

When $\kappa\geq3$ and the energy $E$ is very close to zero, Anderson localizations do not occur for a given potential strength $\lambda$ (refer to Figs. 6 and 7).
If $\lambda$ is extremely large, i.e., $\lambda\rightarrow\pm\infty$, and when $\kappa=3$,  the mobility edge can be determined using Eq. (32), i.e.,
\begin{align}
&E_c=\pm\frac{\sqrt{1-\tau}}{|\lambda|}, \ as \ \lambda\rightarrow\pm\infty .
\end{align}
It has been demonstrated that regardless of the strength of the quasiperiodic hopping parameter $\lambda$, there will always be a range of energies, specifically $-\frac{\sqrt{1-\tau}}{|\lambda|}<E<\frac{\sqrt{1-\tau}}{|\lambda|}$, where critical states exist. As the value of $\lambda$ increases, the energy window for these critical states becomes increasingly smaller.

For the given parameters $\kappa=3$, $\tau=-2$, and $\lambda/t=1.5$, the standard deviations of all the eigenstates are reported in Fig. 8. It is shown that, in comparison with the localized states, the critical states have much larger fluctuations of standard deviations.

\begin{figure}
\begin{center}
\includegraphics[width=1.0\columnwidth]{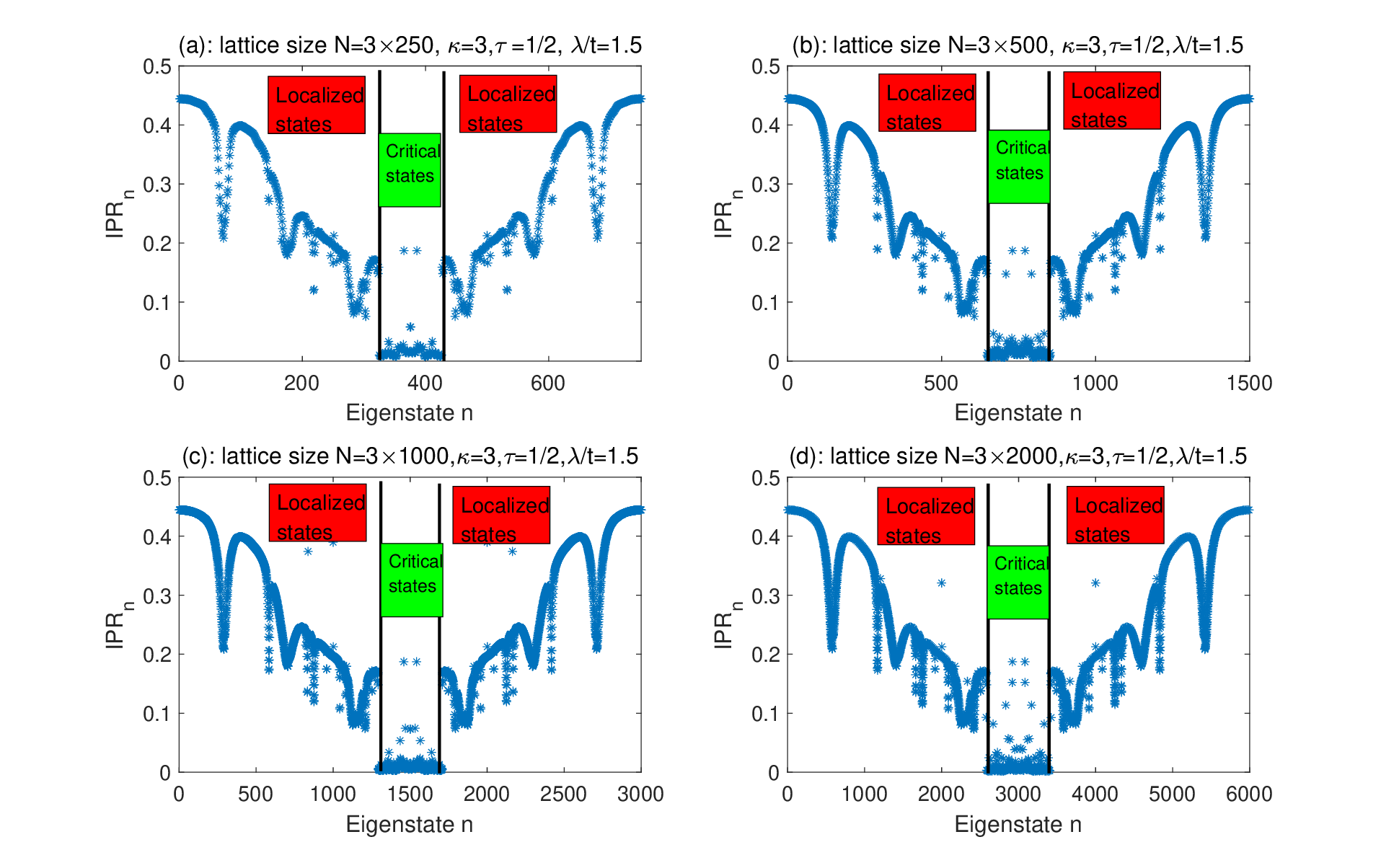}
\end{center}
\caption{The inverse
participation ratio $IPR_n$ of all the eigenstates for system size $N=3\times250,3\times500, 3\times1000$ and $ N=3\times2000$. The eigenenergy $E_n$ increases gradually as the eigenstate index $n$ runs from $1$ to $N$ (along the black dashed line of Fig.7).}
\end{figure}

In order to investigate the properties of  the wave functions of critical states, we also numerically calculate the inverse participation ratio $IPR_n$ of  all eigenstates for different system sizes $N=3\times 250,3\times 500,3\times1000$ and $N=3\times2000$
 \cite{Xiaopeng,Deng}, i.e.,
\begin{align}
&IPR_n=\sum_{i}|\psi_n(i)|^4.
\end{align}
where $\psi_n(i)$ is the normalized wave function for the $n$-th eigenstate. The results are reported in Fig. 9. We find that the IPR of localized states are basically the same for different system sizes $N$, while the IPR of critical states have much larger fluctuations.

Overall, the inverse participation ratio (IPR) of critical states decreases as the system size $N$ increases. This trend can be described by the function
\begin{align}\label{H0}
\overline{IPR}\propto 1/N^x,
\end{align}
where $\overline{IPR}$ is the average value of the IPR within a typical energy interval. It is commonly believed that the scaling exponent $x=0$ for localized states, while for extended states (such as plane wave states), the exponent is $x=1$. For critical states, the exponent should be $0<x<1$.
For different system sizes $N$, due to randomness of $IPR$ of critical states (see Fig.9),  it is difficult to get a definite scaling exponent $x$.
Here we find that for the critical states in the energy interval $-0.178<E/t<0.178$, its average value $\bar{x}\simeq0.47$ (see Fig.11).

\begin{figure}
\begin{center}
\includegraphics[width=1.1\columnwidth]{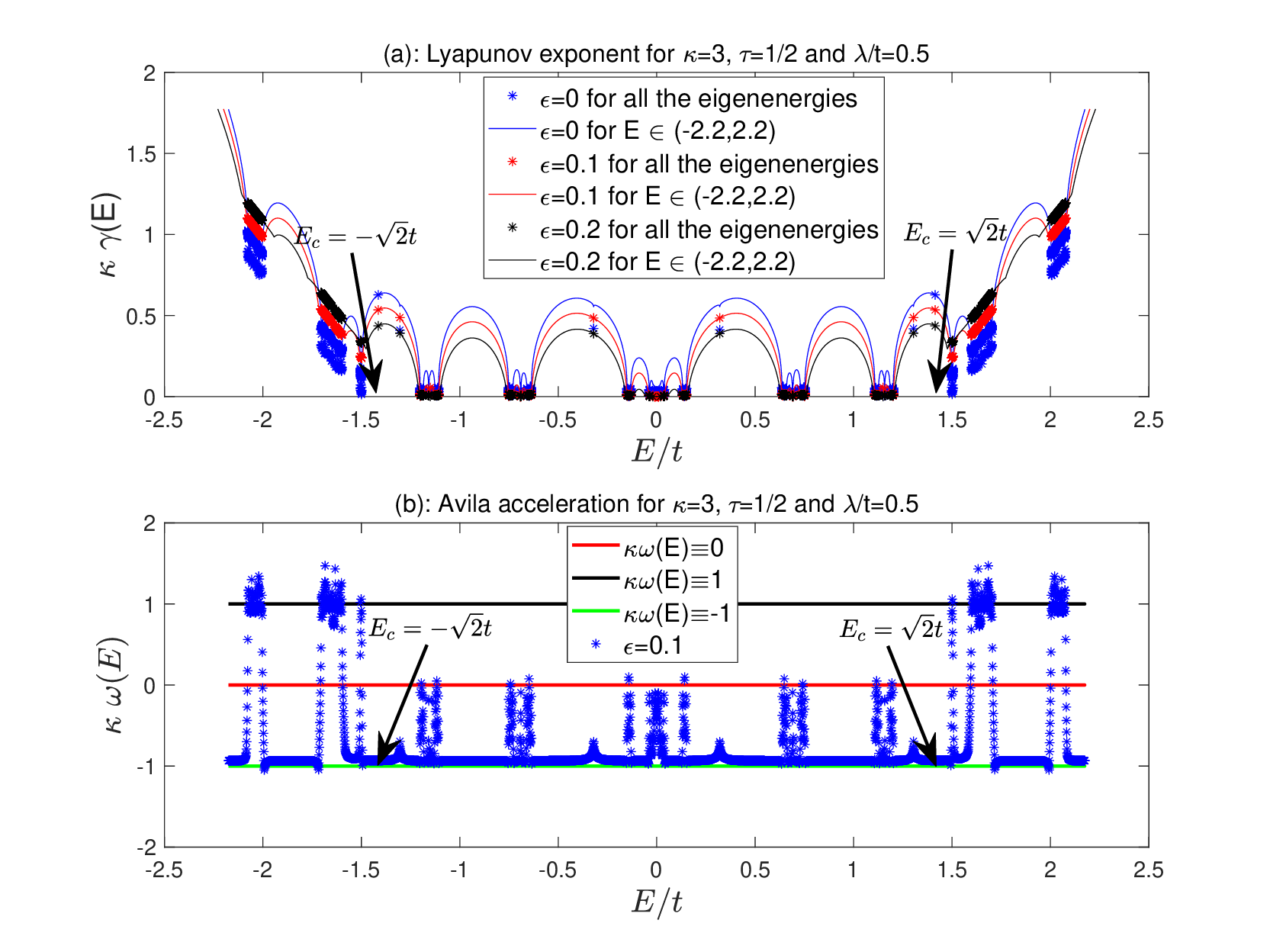}
\end{center}
\caption{ Lyapunov exponents and Avila's accelerations. (a): Lyapunov exponents for $\kappa=3$, $\tau=1/2$ and $\lambda/t=0.5$ (along the blue dashed line of Fig.7).   (b): Avila's acceleration  for for $\kappa=3$, $\tau=1/2$ and $\lambda/t=0.5$. The mobility edges $E_c=\pm\sqrt{2}t$ are indicated by black arrows.  }
\end{figure}

\begin{figure}
\begin{center}
\includegraphics[width=1.1\columnwidth]{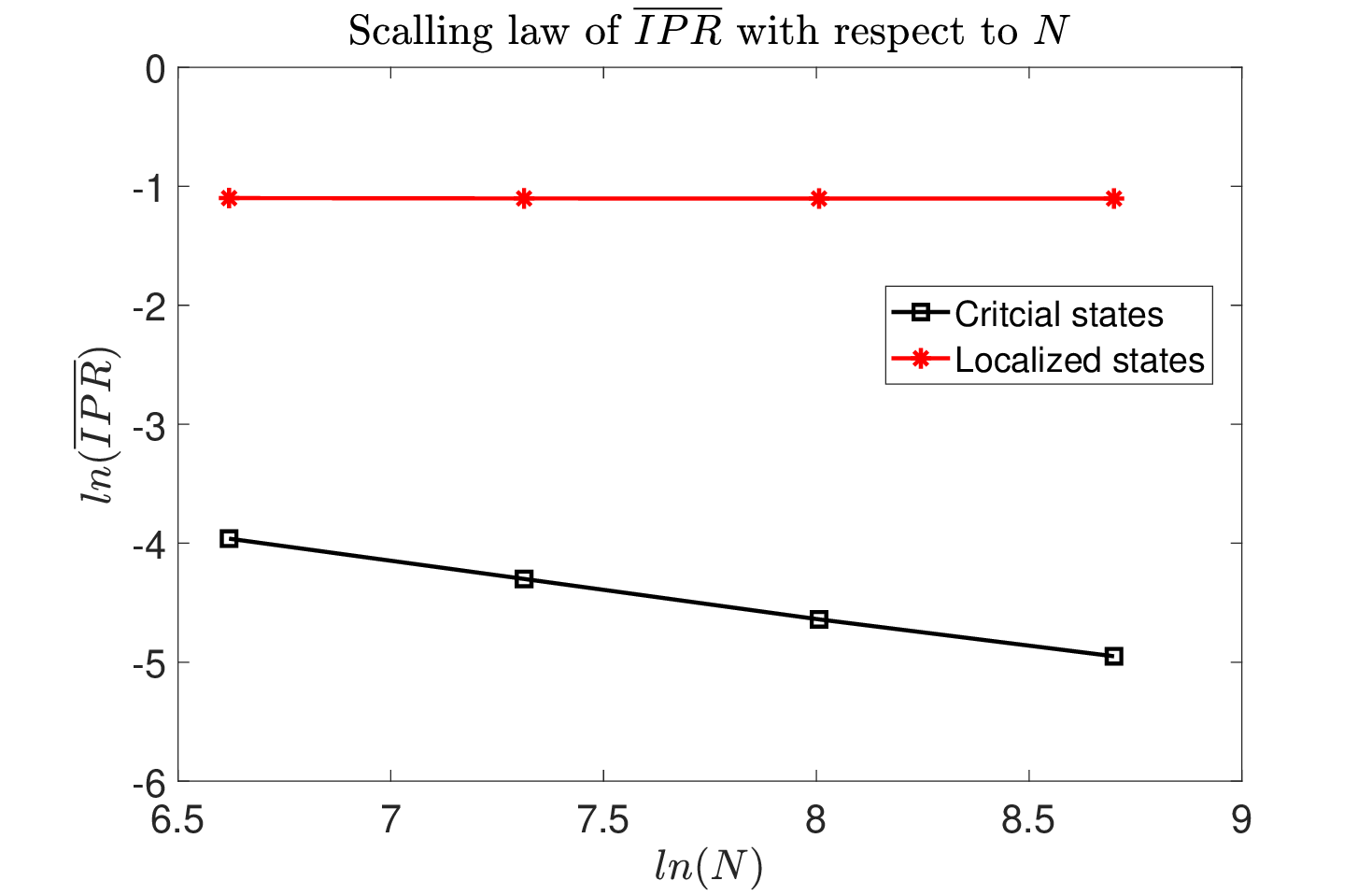}
\end{center}
\caption{ The scaling law of $\overline{IPR}$. The typical energy intervals for localized states and critical states are $-3.63 < E/t < -2.12$,  and $-0.178 <
E/t < 0.178$, respectively. The system sizes for the discrete points
are $N = 3\times250$, $3\times 500$, $3\times 1000$ and $N = 3\times 2000$, respectively. It is found that
for localized states, the average scaling exponent $\bar{x}\simeq0$ and the scaling exponent for critical stats $\bar{x}\simeq0.47$.
 }
\end{figure}

\subsection{Avila acceleration}
In addition, for the bounded quasi-periodic potentials, Avila also defined the acceleration  $\omega(E)$ by \cite{Avila2015}
\begin{align}
\kappa\omega(E)=lim_{\epsilon\rightarrow 0_+}\frac{\gamma(E,\epsilon)-\gamma(E,0)}{\epsilon}.
\end{align}
Using Eqs.(26), when real number $E$ is an eigenvalue,  we get the Avila acceleration
\begin{align}
&\kappa \omega(E)=
\left\{\begin{array}{cccc}
1, \ for  \ energy \ of \ localized \ state \\
0, \ for  \  energy \ of \ critical \ state
  \end{array}\right.
\end{align}
The above results are verified by our numerical calculations as shown in Fig. 10.
Specifically, we have used the values $\tau=1/2$, $\lambda/t=0.5$, and $\epsilon=0,0.1,0.2$ to calculate the Lyapunov exponents using Eq. (13) with the complexified phase $\phi\rightarrow \phi+i\epsilon=i\epsilon$. These calculations were performed for the energy interval $-2.2\leq E\leq2.2$, as shown by the three solid lines in panel (a) of Fig. 10. In our calculations, we have used the parameters $L=200$, $\psi(0)=0$, and $\psi(1)=1$ in Eq. (13). Additionally, we have also calculated the Lyapunov exponents for all eigenenergies using the same parameters, as shown by the three sets of discrete points in panel (a) of Fig. 10. Our results indicate that when $E$ is an eigenenergy of the critical state [$\gamma(E)=0$], the Lyapunov exponents are the same for all three values of $\epsilon=0,0.1,0.2$. However, when $E$ is an eigenenergy of a localized state [$\gamma(E)>0$], the Lyapunov exponents differ for the three values of $\epsilon=0,0.1,0.2$. Furthermore, these differences are linearly proportional to $\Delta\epsilon=0.1$, as shown in Fig. 10.

By taking $\epsilon=0.1$, we also  approximately calculate the Avila's  acceleration $\omega(E)$ by \cite{yicai}
\begin{align}
\kappa\omega(E)\simeq\frac{\gamma(E,\epsilon)-\gamma(E,0)}{\epsilon},
\end{align}
[see panel (b) of Fig.10]. It shows that when $E$ is an eigenenergy of localized state [$\gamma(E)>0$], the Avila's  acceleration is 1. When $E$ is an eigenenergy of the critical state [$\gamma(E)=0$], Avila's acceleration is 0. Additionally, it should be noted that if $E$ is not an eigenenergy, Avila's acceleration is $-1$.

 Further, by combining Eq. (27) and Eq. (43), one can classify systems with different real parameters $E$ (representing different phases) based on their Lyapunov exponent and quantized acceleration, i.e.,
\begin{align}
 &(a): \gamma(E)>0 \ \ \& \ \kappa\omega(E)=-1, \ if\ \ E \ is \ not\ eigenvalue \notag\\
 &(b): \gamma(E)>0 \ \ \& \ \kappa\omega(E)=1,  \ for \ localized \ state\notag\\
 &(c): \gamma(E)=0 \ \ \& \ \kappa\omega(E)=0,  \ for \  critical \ state.
\end{align}

\section{summary}
 In conclusion, we have investigated the localization properties of a one-dimensional lattice model with off-diagonal mosaic quasiperiodic hopping. Our findings show that the parity of the mosaic period has a significant impact on the localization of zero-energy states. 
 Specifically, when the mosaic period is odd, there is always an energy window for critical states, regardless of the strength of the hopping. However, for an even period, the states near zero energy become localized edge states when the hopping strength is sufficiently large.
  In addition,  when mosaic period is even integer, the geometric mean value of the absolute values of energy spectrum is equal to the constant hopping.
 While for an odd mosaic period, the geometric mean value of the energy spectrum is equal to the geometric mean value of the hopping.
  We have also identified mobility edges that separate the localized states from the critical states. Within the critical region, we have observed large fluctuations in the spatial extensions of the eigenstates.

 Our results, obtained using Avila's theory, reveal the exact values of the Lyapunov exponents and mobility edges. Additionally, we have determined that the critical index of the localization length is $\nu=1$. Our numerical results show that the scaling exponent of the inverse participation ratio (IPR) for critical states is approximately $x\simeq0.47$, confirming that these states are indeed critical. Furthermore, we have demonstrated that the Lyapunov exponent and Avila's acceleration can be used to classify systems with different values of $E$.

\section*{Acknowledgements}
This work was supported by the NSFC under Grants Nos.
11874127, 12401208,  12171039, 12375014, the  Natural Science Foundation of Jiangsu Province (Grants No BK20241431), Heilongjiang Provincial Natural Science Foundation of
China with No. LH2019015, the Joint Fund with
Guangzhou Municipality under No.
202201020137, and the Starting Research Fund from
Guangzhou University under Grant No.
RQ 2020083.


\end{document}